\documentclass[english]{article}

\usepackage[colorlinks,bookmarksopen,bookmarksnumbered,allcolors=red]{hyperref}
\usepackage{xcolor}
\usepackage{graphicx}
\usepackage{geometry}
\usepackage{amsmath, amssymb, bm}
\usepackage{authblk}
\usepackage{placeins}
\usepackage{pdflscape}
\usepackage{rotating}
\usepackage{multirow}

\definecolor{bluegreen}{RGB}{46,141,131}
\definecolor{darkgreen}{RGB}{46,139,87}
\definecolor{darkred}{RGB}{219,7,61}
\definecolor{darkblue}{RGB}{0,0,137}

\title{\textbf{Approximate Bayesian inference for joint partially linear modeling of longitudinal measurements and spatial time-to-event data}}

\author[,1]{T. Baghfalaki\thanks{Corresponding author: taban.baghfalaki@manchester.ac.uk}}
\author[2]{M. Ganjali}
\author[3]{R. Martins}
\affil[1]{Department of Mathematics, The University of Manchester, Manchester, UK}
\affil[2]{Department of Statistics, Faculty of Mathematical Sciences, Shahid Beheshti University, Tehran, Iran.}
\affil[3]{Faculdade de Ciências da Universidade de Lisboa (FCUL), Centro de Estatística e Aplicações da Universidade de Lisboa (CEAUL), Lisboa, Portugal.}

\date{}

\begin{document}

\maketitle

\begin{abstract}
The integration of longitudinal measurements and survival time in statistical modeling offers a powerful framework for capturing the interplay between these two essential outcomes, particularly when they exhibit associations. However, in scenarios where spatial dependencies among entities are present due to geographic regions, traditional approaches may fall short. In response, this paper introduces a novel approximate Bayesian hierarchical model tailored for jointly analyzing longitudinal and spatial survival outcomes. The model leverages a conditional autoregressive structure to incorporate spatial effects, while simultaneously employing a joint partially linear model to capture the nonlinear influence of time on longitudinal responses. Through extensive simulation studies, the efficacy of the proposed method is rigorously evaluated. Furthermore, its practical utility is demonstrated through an application to real-world HIV/AIDS data sourced from various Brazilian states, showcasing its adaptability and relevance in epidemiological research.\\
{Keywords:} Approximate Bayesian inference; Integrated Laplace approximation (INLA); Joint modeling; Latent Gaussian model; Spatial effects; Spline functions.
\end{abstract}

\section{Introduction}
In numerous clinical longitudinal studies, data on longitudinal markers are gathered until an event occurs \cite{rizopoulos2012joint,elashoff2016joint}. For instance, in HIV studies, CD4 counts are typically monitored until a patient's demise or dropout from the study, highlighting the interplay between longitudinal and survival processes. However, these processes are often analyzed separately, leading to biased parameter estimates \cite{guo2004separate}. The solution lies in joint modeling, a burgeoning field with extensive medical applications \cite{chen2014joint,temesgen2018joint,loureiro2019modelling,chesnaye2020introduction,gebrerufael2020statistical,tesfay2021joint}.\\
Joint modeling typically involves two submodels: the longitudinal submodel, often a linear mixed effects model \cite{wu2009mixed}, and the survival submodel, ranging from the traditional Cox model to accelerated failure time models \cite{legrand2021advanced}. Central to joint modeling is the establishment of a linking structure, often achieved by sharing random effects between the longitudinal and survival processes \cite{rizopoulos2012joint,elashoff2016joint}.\\
Recent years have seen significant advancements in research within this domain, extending beyond the conventional framework to analyze diverse data scales such as counts or ordinal longitudinal outcomes \cite{li2010joint,armero2016bayesian,baghfalaki2020bayesian,zhu2018joint,baghfalaki2021approximate}, as well as multivariate longitudinal outcomes \cite{chi2006joint,baghfalaki2014joint,luo2014bayesian,he2016joint,kang2022joint}. Additional extensions include addressing competing risks, multiple causes of death \cite{williamson2008joint,li2010joint,hickey2018joint,teixeira2019joint,baghfalaki2020bayesian}, and cure fraction models \cite{chen2004new,yang2021joint}.
For a comprehensive review of joint modeling, readers may refer to \cite{sousa2011review,papageorgiou2019overview,furgal2019review,alsefri2020bayesian}.\\
A promising direction for further research is the incorporation of spatial effects based on geographical regions, an area that has been relatively underexplored \cite{martins2016bayesian,martins2017joint,momenyan2021joint}. 
Although joint modeling is highly effective, challenges remain in parameter estimation, which frequently involve complex and time-consuming computations. In the realm of joint modeling, various strategies for parameter estimation are employed. Classical approaches, such as the EM algorithm \cite{hsieh2006joint,ganjali2015copula,ha2017h,mehdizadeh2021two,murray2022fast}, have been widely utilized alongside Bayesian methods \cite{alsefri2020bayesian}. However, these methods often involve intricate and time-consuming computations, prompting the exploration of alternative strategies. 
To address this, researchers have explored approximate Bayesian methods as an efficient alternative to traditional approaches \cite{van2019new,baghfalaki2021approximate,niekerk2021competing,rustand2022fast}. \\
{Approximate Bayesian methods are often employed when exact Bayesian inference becomes computationally prohibitive due to the complexity or dimensionality of the model. Several approaches exist to address these challenges. {Markov Chain Monte Carlo (MCMC)} methods are a widely used class of algorithms that generate samples from the posterior distribution. However, while flexible, they can be computationally intensive, especially in high-dimensional settings \cite{gilks1995}. {Approximate Bayesian Computation (ABC)} is another approach designed for models with intractable likelihoods, where inference is performed by comparing observed data to data simulated under the model \cite{beaumont2010,sisson2018}. ABC is often applied in fields such as genetics and ecology, where exact likelihood functions are unavailable.\\
Alternatively, methods such as {Variational Inference (VI)} and {Integrated Nested Laplace Approximation (\texttt{INLA})} offer more computationally efficient solutions. VI approximates the posterior by finding a simpler distribution that minimizes the divergence from the true posterior, often at the cost of some accuracy \cite{blei2017}. \texttt{INLA}, on the other hand, is particularly effective for latent Gaussian models, using Laplace approximations to provide fast and accurate approximations of marginal posteriors \cite{rue2009,rue2017}.}\\
In this paper, we propose an approximate Bayesian method for joint modeling of longitudinal and spatial survival data. Our approach builds upon existing models while introducing innovations to enhance computational efficiency and model flexibility. Performance evaluation through simulation studies and application to real-world HIV/AIDS data from Brazil underscores the utility of our method. For implementation, we leverage the \texttt{INLA} package, which allows for efficient Bayesian inference using Integrated Nested Laplace Approximations. The \texttt{INLA} framework is particularly well-suited for high-dimensional latent models like those involving spatial and longitudinal data. Additionally, we provide accompanying R code to ensure reproducibility and facilitate the application of our method in other research settings. By combining the strengths of \texttt{INLA} with our methodological enhancements, our approach offers a robust and computationally feasible solution to the challenges of joint modeling in complex data scenarios.\\
%In this paper, we propose an approximate Bayesian method for joint modeling of longitudinal and spatial survival data. Our approach builds upon existing models while introducing innovations to enhance computational efficiency and model flexibility. Performance evaluation through simulation studies and application to real-world HIV/AIDS data from Brazil underscores the utility of our method. For implementation, we leverage the \texttt{INLA} package, with accompanying R code provided for reproducibility.\\
We employ an approximate Bayesian approach to implement joint modeling of longitudinal and spatial survival data. While the core model resembles that of \cite{martins2017joint}, we introduce two key innovations: (1) leveraging approximate Bayesian inference instead of the traditional MCMC method, resulting in reduced computational time compared to Gibbs sampling, and (2) incorporating a partial linear joint model to account for nonlinear effects of time, thereby enhancing model versatility. To evaluate the performance of our proposed model, we conduct simulation studies and apply it to real-world Brazilian HIV/AIDS data. We utilize the \texttt{INLA} package for implementation, and the accompanying R code is publicly available at \url{https://github.com/tbaghfalaki/ASJM}.\\
The structure of this paper is organized as follows: Section 2 details the specification of the joint modeling framework, including longitudinal submodels, spatial survival submodel, and the approximate Bayesian approach for model implementation. Section 3 
focuses on the analysis of real data, while Section 4  
presents the results of simulation studies. Finally, Section 5 provides concluding remarks and discusses avenues for future research.

\section{Joint modeling specification}
\subsection{Notation}
The spatial joint model has two separate submodels including a linear mixed effects model and a spatial survival model. The association of these two separate models are taken into account by a shared random effect. Let $N$ individuals be collected from $K$ regions such that the number of individuals in each region is $n_k$ and $\sum_{k=1}^K n_k=N$.\\
Let $T_{ik}^*$ be the true value of survival time for the $i^{th}$ individual in the $k^{th}$ region, such that the observed survival time $T_{ik}$ is the minimum value of $T_{ik}^*$ and the censoring time $C_{ik}$, that is, $T_{ik}=min(T_{ik}^*,C_{ik})$. Consider an indicator $\delta_{ik}$ such that  $\delta_{ik}=0$ indicates a right censored observation. Thus, the survival outcome is the pair $(T_{ik},\delta_{ik}),~i=1,\cdots,n_k,k=1,\cdots,K$. \\
The longitudinal outcomes for each individual are collected at time points $s_{ikj}<T_{ik},~i=1,\cdots,n_k,k=1,\cdots,K,j=1,\cdots,m_{ik}$, where $m_{ik}$ is the number of repeated measurements for the $i^{th}$ individual in the $k^{th}$ region. Therefore, the observed vector of longitudinal outcome is given by $\bm{y}_{ik}=(y_{ik1},\cdots,y_{ikm_{ik}})^\prime$, where $y_{ikj}=y_{ik}(s_{ikj})$ and it is obtained as the sum of the true value of the longitudinal outcome at the same time point, $y_{ik}^*(s_{ikj})$, and an error term, that is, $y_{ik}(s_{ikj})=y_{ik}^*(s_{ikj})+\epsilon_{ik}(s_{ikj})$.
\subsection{Longitudinal submodel}
We consider a semi-parametric mixed effects model for describing the longitudinal process as follows \begin{eqnarray}
y_{ik}(s_{ikj})=y_{ik}^*(s_{ikj})+\epsilon_{ik}(s_{ikj})=\bm{x}_{1ik}^\prime(s_{ikj}) \bm{\beta}+g(s_{ikj})+\bm{z}_{ik}^\prime(s_{ikj}) \bm{b}_{ik}+\epsilon_{ik}(s_{ikj}), \end{eqnarray}
where $\bm{x}_{1ik}(s)$ is a $p_1$-dimensional vector of explanatory variables at time point $s$ corresponding to the regression coefficients $\bm{\beta}$,  $\bm{z}_{ik} (s) $ denotes a $q$-dimensional vector of explanatory variables at time point $s$ corresponding to the random effects $\bm{b}_{ik}$, $\bm{b}_{ik}\sim N_q(\bm{0},\bm{D})$, $\epsilon_{ik}(s)$s are the error terms which are assumed to be independently and identically distributed as $N(0,\sigma^2)$ and $g(s)$ is a nonlinear function. The shape of $g(.)$  is typically unknown and it should be estimated using some  semi-parametric approaches such as spline functions.\\
The spline function with fixed
knots sequence $\kappa_{1}<\cdots<\kappa_{l}$
and fixed degree $d$ can be given by:
\begin{eqnarray}%\label{e1}
{g}(s) = \sum\limits_{j = 1}^{l + d + 1} {{\alpha^*_{j}}{B_j}(s)},
\end{eqnarray}
where $B_j$s
are basis functions defining a vector space and $\alpha^*_j$s are the associated spline coefficients. 
Some of the possible basis functions are  the truncated power basis, the B-spline basis and the natural spline. For more details see
\cite{wu2006nonparametric} and \cite{baghfalaki2021approximate}.\\
The number  and the  positions of the knots are important issues of
the spline function. The use of a penalized spline is a popular strategy for this purpose \cite{wu2006nonparametric}. 
Another strategy for this aim is  to determine the optimal number of knots by using, for instance,  generalized cross validation, AIC, etc. \cite{eilers1996flexible,wu2006nonparametric}. 
For considering the  spline function \texttt{INLA} in sofware R  mentioning the knots, giving the degree and adding the basis functions as some new covariates to the model are necessary \cite{wang2018bayesian}. 
\subsection{Spatial survival submodel}
For the survival outcome, a proportional hazard model considered as follows:
\begin{eqnarray}\label{zi3}
{h_{ik}}(t) = {h_{0i}}(t)\exp \left\{
{\bm{x}_{2ik}}(t)^\prime{\bm{\alpha} } +\bm{\gamma}^\prime \bm{b}_{ik}+\nu_k
\right\},
\end{eqnarray}
where $\bm{x}_{2ik}(t)$ is the $p_2$-dimensional vector of explanatory variables, 
  $\bm{\alpha} $ is the fixed effects corresponding to  ${\bm{x}_{2ik}}({t})$, 
  $\bm{\gamma}$ denotes a vector of the parameters for  measuring the association between two models, ${h_{0i}}(t)=\varphi t^{\varphi -1},~\varphi>0$ which leads to Weibull hazard function and $\nu_k$, for ~$k=1,\cdots,K$ represent spatial  random effects. 
  Although, we consider only the Weibull  hazard function,  because of its simplicity,
   other forms might be considered as well.\\
%  A proper non-intrinsic Besag model \cite{besag1974spatial,besag1975statistical} is applied to consider   such that the conditional distribution of each areal unit given the other units is defined as:
%\begin{eqnarray}\label{car10}
%\nu_k|\bm{\nu}_{-k}\sim N(\frac{1}{n_k}\sum_{k\sim k'} \nu_k,\frac{1}{\tau n_k}),
%\end{eqnarray}
%where $n_k$ is the number of neighbours of region $k$, $\bm{\nu}=(\nu_1,\cdots,\nu_K)^\prime$, 
%$ k\sim k'$ indicates that the two regions $k$ and $k'$ are neighbours, $\bm{\nu}_{-k}=(\nu_1,\cdots,\nu_{k-1},\nu_{k+1},\cdots,\nu_K)^\prime$ and $\tau>0$ is a scaling parameter.  By considering the Brook’s lemma \cite{brook1964distinction}, given the $K$ normal full conditional \eqref{car10}, $\bm{\nu}$ is a Gaussian Markov random field \cite[GMRF]{rue2009approximate,rue2005gaussian} with mean $\bm{0}$ and precision matrix $\tau \bm{\Omega}$, being its entries defined as $\omega_{kk}=n_k$ and $\omega_{kj}=1,~k\neq j$. We can use  the following notation instead: 
%\begin{eqnarray}\label{lll2}
%\bm{\nu}\sim N_K (\bm{0},\tau^{-1} %\bm{\Omega}^{-1}).
%\end{eqnarray}
%In this paper, we use the second proper version of the Besag model \cite{rue2009approximate} as distributional assumption for $\bm{\nu}$ which is corrected by a ratio $\zeta$ ($\in (0,1)$) as follows: 
%\begin{eqnarray}\label{lll3}
%\bm{\nu}\sim N_K (\bm{0},\tau^{-1}((1-\zeta) \bm{I}_K+\zeta \bm{\Omega}^{-1})),
%\end{eqnarray}
%where $\bm{I}_K$ is a $K \times K$ identity matrix. 
{A proper non-intrinsic Besag model \cite{besag1974spatial,besag1975statistical} is applied to consider such that the conditional distribution of each areal unit given the other units is defined as:
\begin{eqnarray}\label{car10}
\nu_k|\bm{\nu}_{-k}\sim N\left(\frac{1}{n_k}\sum_{k\sim k'} \nu_k,\frac{1}{\tau n_k}\right),
\end{eqnarray}

where \(n_k\) is the number of neighbors of region \(k\), \(\bm{\nu} = (\nu_1, \cdots, \nu_K)'\), \(k \sim k'\) indicates that the two regions \(k\) and \(k'\) are neighbors (i.e., they share a boundary or are adjacent to each other), \(\bm{\nu}_{-k} = (\nu_1, \cdots, \nu_{k-1}, \nu_{k+1}, \cdots, \nu_K)'\) represents the vector of values from all regions except region \(k\), and \(\tau > 0\) is a scaling parameter, also referred to as the precision parameter.

This equation describes the conditional distribution of \(\nu_k\) (the value of the spatial variable in region \(k\)) given all other values \(\bm{\nu}_{-k}\). It follows a normal distribution where the mean is the average value of the neighboring regions \(k'\), weighted by the number of neighbors \(n_k\). The variance of the conditional distribution is inversely proportional to both the number of neighbors \(n_k\) and the precision parameter \(\tau\).

This setup implies that the value of \(\nu_k\) depends primarily on the values of its neighboring regions and suggests that areas with more neighbors will have a more tightly constrained (lower variance) value for \(\nu_k\). This conditional distribution is a key element of the Gaussian Markov Random Field (GMRF) structure, where each region is conditionally dependent on its neighbors.

By considering Brook's lemma \cite{brook1964distinction}, given the \(K\) normal full conditional \eqref{car10}, \(\bm{\nu}\) is a Gaussian Markov random field \cite[GMRF]{rue2009approximate,rue2005gaussian} with mean $\bm{0}$ and precision matrix \(\tau \bm{\Omega}\), being its entries defined as \(\omega_{kk} = n_k\) and \(\omega_{kj} = 1\), \(k \neq j\). We can use the following notation instead:
\begin{eqnarray}\label{lll2}
\bm{\nu}\sim N_K (\bm{0},\tau^{-1} \bm{\Omega}^{-1}).
\end{eqnarray}
Here, \(\bm{\nu} = (\nu_1, \dots, \nu_K)'\) is the vector representing the spatial variable in each of the \(K\) regions. The notation \(\bm{\nu} \sim \mathcal{N}_K(0, \tau^{-1} \bm{\Omega}^{-1})\) indicates that \(\bm{\nu}\) follows a multivariate normal distribution with mean 0 and a precision matrix \(\tau \bm{\Omega}\).

- \(\tau^{-1}\) is the scaling parameter (the inverse of the precision parameter \(\tau\)).\\
- \(\bm{\Omega}\) is the precision matrix, which encodes the spatial dependencies between regions. The structure of \(\bm{\Omega}\) is based on the neighborhood system of the regions:\\
  - \(\omega_{kk} = n_k\), the number of neighbors for region \(k\),\\
  - \(\omega_{kj} = 1\) if regions \(k\) and \(j\) are neighbors,\\
  - \(\omega_{kj} = 0\) if regions \(k\) and \(j\) are not neighbors.
  
This precision matrix \(\bm{\Omega}\) ensures that the spatial dependencies are captured properly, as it reflects the strength of the relationships between neighboring regions. A high precision value means stronger dependency, while regions that are not neighbors will have no direct dependency (i.e., \(\omega_{kj} = 0\)).

In this paper, we use the second proper version of the Besag model \cite{rue2009approximate} as the distributional assumption for \(\bm{\nu}\) which is corrected by a ratio \(\zeta \in (0,1)\) as follows:
\begin{eqnarray}\label{lll3}
\bm{\nu}\sim N_K (\bm{0},\tau^{-1}((1-\zeta) \bm{I}_K+\zeta \bm{\Omega}^{-1})),
\end{eqnarray}
where $\bm{I}_K$ is a $K \times K$ identity matrix. 

This equation introduces a correction to the standard Besag model. The parameter \(\zeta\) is used to adjust the influence of the spatial structure (encoded in \(\bm{\Omega}\)) versus independent variability (encoded in \(\bm{I}_K\), the identity matrix).

- \(\bm{I}_K\) is the identity matrix of size \(K \times K\), which represents independent variation for each region. If only \(\bm{I}_K\) were present (i.e., \(\zeta = 0\)), it would imply no spatial dependence, and each region would have an independent normal distribution with variance \(\tau^{-1}\).\\
- \(\bm{\Omega}^{-1}\) represents the spatial structure, where \(\bm{\Omega}\) is the precision matrix derived from the neighborhood relationships of the regions.\\
- The term \((1-\zeta) \bm{I}_K + \zeta \bm{\Omega}^{-1}\) is a weighted combination of the identity matrix and the inverse precision matrix, allowing a balance between independent variation and spatial dependence. The parameter \(\zeta\) controls this balance:\\
  - When \(\zeta = 1\), the model fully depends on the spatial structure (as in the original Besag model).\\
  - When \(\zeta = 0\), the model implies no spatial correlation, and the regions are modeled independently.

This modification is essential to ensure the model is proper (i.e., the covariance matrix is positive definite) and that the spatial structure is still retained, but in a controlled manner. The model remains flexible, as \(\zeta\) can be tuned to allow for different levels of spatial dependence in the data.
}

\subsection{Approximate Bayesian approach}
In this section, we use the approximate Bayesian approach to perform inference for the spatial joint model. For this aim, at the first stage 
we specify the spatial joint model in term of latent Gaussian models (LGM), then we apply integrated nested  Laplace approximation (\texttt{INLA}) to approximate the joint posterior density of the model.  \\
{\textbf{Spatial joint model in term of latent Gaussian models}}\\
%Let $\eta_{ikj}^y=\bm{x}_{1ik}^\prime(s_{ikj}) \bm{\beta}+g(s_{ikj})+\bm{z}_{ik}^\prime(s_{ikj}) \bm{b}_{ik}$, $\eta_{ik}^t=\bm{x}_{2ik}(t)^\prime{\bm{\alpha} } +\bm{\gamma}^\prime \bm{b}_i+\nu_k$, $\bm{\eta}_{ik}^y=(\eta_{i11}^y,\cdots,\eta_{ikm_{ik}}^y)$,
%$\bm{\eta}_{k}^y=(\eta_{1k}^y,\cdots,\eta_{n_k,k}^y)$,
%$\bm{\eta}_k^t=(\eta_{1k}^t,\cdots,\eta_{n_k,k}^t)$ and $\bm{\eta}_k=(\bm{\eta}_k^y,\bm{\eta}_k^t)$. Also, $\bm{y}=(\bm{y}_{11},\cdots,\bm{y}_{n_KK})$, 
%$\bm{T}=({T}_{11},\cdots,T_{n_KK})$, $\bm{\delta}=({\delta}_{11},\cdots,\delta_{n_KK})$, $\bm{x}_1=(\bm{x}_{111},\cdots,\bm{x}_{1n_KK})$, $\bm{x}_2=(\bm{x}_{211},\cdots,\bm{x}_{2n_KK})$ and $\bm{z}=(\bm{z}_{11},\cdots,\bm{z}_{n_KK})$. Also, let $\bm{\theta}_1$ be the vector of hyperparameters  and $\bm{\Lambda}=\{\bm{\eta}_{1},\cdots,\bm{\eta}_{K},\bm{\beta}, \bm{\alpha},\bm{\gamma}\}$ be the latent structure and it is considered to be  GMRF.
{Let \(\eta_{ikj}^y\) denote the latent variable for the longitudinal outcome of the \(i^{th}\) individual in the \(k^{th}\) region at time \(s_{ikj}\), defined as 
\[
\eta_{ikj}^y = \bm{x}_{1ik}^\prime(s_{ikj}) \bm{\beta} + g(s_{ikj}) + \bm{z}_{ik}^\prime(s_{ikj}) \bm{b}_{ik}.
\]
The latent variable for the survival outcome of the same individual is 
\[
\eta_{ik}^t = \bm{x}_{2ik}(t)^\prime \bm{\alpha} + \bm{\gamma}^\prime \bm{b}_i + \nu_k.
\]
The vectors \(\bm{\eta}_{ik}^y = (\eta_{i11}^y, \ldots, \eta_{ikm_{ik}}^y)\) and \(\bm{\eta}_{k}^y = (\eta_{1k}^y, \ldots, \eta_{n_k,k}^y)\) summarize the longitudinal latent variables for individuals and regions, while \(\bm{\eta}_k^t = (\eta_{1k}^t, \ldots, \eta_{n_k,k}^t)\) captures the survival latent variables for the \(k^{th}\) region. The overall observed longitudinal outcomes are represented by $\bm{y} = (\bm{y}_{11}, \ldots, \bm{y}_{n_KK}),$
the observed survival times by 
$
\bm{T} = (T_{11}, \ldots, T_{n_KK}),
$
and the censoring indicators by 
$
\bm{\delta} = (\delta_{11}, \ldots, \delta_{n_KK}).
$
The covariate matrices are defined as 
$
\bm{x}_1 = (\bm{x}_{111}, \ldots, \bm{x}_{1n_KK})
$
for longitudinal outcomes and 
$
\bm{x}_2 = (\bm{x}_{211}, \ldots, \bm{x}_{2n_KK})
$
for survival analysis, while 
$
\bm{z} = (\bm{z}_{11}, \ldots, \bm{z}_{n_KK})
$
contains random effects covariates. Additionally, let \(\bm{\theta}_1\) be the vector of hyperparameters, and 
$
\bm{\Lambda} = \{\bm{\eta}_{1}, \ldots, \bm{\eta}_{K}, \bm{\beta}, \bm{\alpha}, \bm{\gamma}\}
$
represent the latent structure, which is considered a GMRF.
}
 $\bm{\eta}$ is specified as a latent
Gaussian random field with the following density function:
 \begin{eqnarray}\label{b27}
\bm{\eta}|\bm{\theta}_1\sim N(\bm{0}, Q^{-1}(\bm{\theta}_1))
 \end{eqnarray}
where $Q(\bm{\theta}_1)$ is a sparse precision matrix with a vector of parameters $\bm{\theta}_2$.\\
The joint distribution (likelihood function) of $\bm{y},\bm{T}$ and $\bm{\delta}$
given $\bm{\Delta},\bm{\theta}_2,\bm{x}_1,\bm{x}_2$ and $\bm{z}$ 
is as follows
\begin{eqnarray}\label{ii}
\pi(\bm{y},\bm{T},\bm{\delta}|\bm{\Delta},\bm{\theta}_2,\bm{x}_1,\bm{x}_2,\bm{z})&=&\prod_{k=1}^{K}
\prod_{i=1}^{n_k}\prod_{j=1}^{m_{ik}} \phi(y_{ikj};\eta_{ikj}^y,\sigma^2)\\\nonumber
&\times& \prod_{k=1}^{K}
\prod_{i=1}^{n_k} {h_{ik}}(t)^{\delta_{ik}} \exp\{-\int_0^{T_{ik}} h_{ik}(s) ds\}.
\end{eqnarray}
where $\phi(.;\mu ,\sigma^2)$ denotes the normal distribution with mean $\mu$ and variance $\sigma^2$,
$h_{ik}(t)=h_{0i}(t)\exp(\eta_{ik}^t)$ is the hazard function given by \eqref{zi3} and $\bm{\theta}_1=\{\sigma^2,\varphi \}$. \\
Note that this likelihood is obtained by the assumption of the joint modeling: 
given the random effects $\bm{b}_{ik}$, the random variables $\bm{y}_{ik}$ and $(T_{ik},\delta_{ik})$ are independent. Also, given $\bm{b}_{ik}$, the components of $y_{ik1},\cdots,y_{ikm_{ik}}$ are independent.\\
Consider a prior $\pi(\bm{\theta})$  for 
$\bm{\theta}=(\bm{\theta}_1,\bm{\theta}_2)$. The joint posterior distribution of $\bm{\Lambda}$ and $\bm{\theta}$ given $\bm{y},\bm{t}$ and $\bm{\delta}$ is as follows:
    \begin{eqnarray}\label{b2}
 \pi (\bm{\Lambda},\bm{\theta} |\bm{y},\bm{t},\bm{\delta} ) \propto \pi(\bm{y},\bm{T},\bm{\delta}|\bm{\Delta},\bm{\theta}_1,\bm{x}_1,\bm{x}_2,\bm{z})
 \pi (\bm{\Lambda}|\bm{\theta}_2  )\pi (\bm{\theta} ).
 \end{eqnarray}
The strategy of \texttt{INLA} is to estimate the marginal distributions of the latent effects and the hyperparameters. For  a latent parameter $\Lambda_j$, we have
 \begin{eqnarray}\label{e1}
 \pi ({\Lambda}_j|\bm{y},\bm{t},\bm{\delta} ) =\int 
 \pi ({\Lambda}_j|\bm{\theta},\bm{y},\bm{t},\bm{\delta} ) \pi(\bm{\theta}|\bm{y},\bm{t},\bm{\delta} ) d\bm{\theta}.
 \end{eqnarray}
Also, the posterior marginal for a hyperparameter $\theta_k$ is as follows:
 \begin{eqnarray}\label{e2}
 \pi ({\theta}_j|\bm{y},\bm{t},\bm{\delta} ) =\int  \pi(\bm{\theta}|\bm{y},\bm{t},\bm{\delta} ) d\bm{\theta}_{-j}.
 \end{eqnarray}
where $\bm{\theta}_{-j}$ is a vector of hyperparameters $\bm{\theta}$ without $\theta_j$.\\
Based on the approximation of \cite{rue2009approximate}, the joint posterior distribution of $\bm{\theta}$, $\tilde{\pi}({\bm{\theta}}|\bm{y},\bm{t},\bm{\delta})$, can be used to compute equations \eqref{e1} and \eqref{e2} as follows:
 \begin{eqnarray}\label{e3}
 \tilde{\pi}({\bm{\theta}}|\bm{y},\bm{t},\bm{\delta})\propto \frac{{\pi}({\bm{\Lambda}},\bm{\theta},\bm{y},\bm{t},\bm{\delta})}{{\tilde{\pi}_G}({\bm{\Lambda}|\bm{\theta}},\bm{y},\bm{t},\bm{\delta})}\Bigg|_{\bm{\Lambda}=\bm{\Lambda}^*(\bm{\theta})},
 \end{eqnarray}
where ${\tilde{\pi}_G}({\bm{\Lambda}|\bm{\theta}},\bm{y},\bm{t},\bm{\delta})$
is a Gaussian approximation of ${{\pi}_G}({\bm{\Lambda}|\bm{\theta}},\bm{y},\bm{t},\bm{\delta})$ and 
$\bm{\Lambda}^*(\bm{\theta})$ is the mode of this distribution for a given value of $\bm{\theta}$.
The posterior marginal distribution of $\pi(\theta_j|\bm{y},\bm{t},\bm{\delta})$ can be approximated by integrating $\bm{\theta}_{-j}$ out in equation \eqref{e3}.

{\textbf{{Approximations for $\pi(\Lambda_j|\bm{\theta},\bm{y},\bm{t},\bm{\delta})$ }}\\
The approximation of $\Lambda_j$ given $\bm{y},\bm{t},\bm{\delta}$ in \texttt{INLA} can be obtained by the following expression:
\begin{eqnarray}\label{e63}
\pi ({\Lambda _j}|\bm{y},\bm{t},\bm{\delta} ) \simeq \sum\limits_{r = 1}^R {\tilde \pi ({\Lambda _j}|{\bm{\theta}^{(r)}},\bm{y},\bm{t},\bm{\delta})} \tilde \pi ({\bm{\theta} ^{(r)}}|\bm{y},\bm{t},\bm{\delta} ){\Delta _r},
 \end{eqnarray}
where $\{\bm{\theta}^{(1)},\cdots,\bm{\theta}^{(R)}\}$ is a set of value of $\bm{\theta}$ for 
computing the numerical integration with weights  $\{\Delta _1,\cdots,\Delta_R \}$.\\
The three different following approximations are described by \cite{rue2009approximate} to approximate 
$\pi ({\Lambda _j}|{\bm{\theta}},\bm{y},\bm{t},\bm{\delta})$:
\begin{enumerate}
  \item Marginalization of the denominator of equation \eqref{e3}.
  \item The use of the Laplace approximation as follows: 
  $${\pi _{LA}}({\Lambda _j}|\theta ,y,t,\delta ) \propto {\left. {\frac{{\pi (\Lambda ,\theta ,y,t,\delta )}}{{{\pi _{G}}({\Lambda _{ - j}}|{\Lambda _j},\theta ,y,t,\delta )}}} \right|_{{\Lambda _{ - j}} = \Lambda _{ - j}^*({\Lambda _j},\theta )}},$$
  where ${{\pi _{G}}({\Lambda _{ - j}}|{\Lambda _j},\theta ,y,t,\delta )}$ is the Gaussian approximation for  ${{\pi }({\Lambda _{ - j}}|{\Lambda _j},\theta ,y,t,\delta )}$ and ${{\Lambda _{ - j}} = \Lambda _{ - j}^*({\Lambda _j},\theta )}$ is its mode. This methodology is computationally expensive, thus, the following approximation is proposed: 
  $${\pi _{LA}}({\Lambda _j}|{\bm{\theta}},\bm{y},\bm{t},\bm{\delta} ) \propto \phi(\Lambda _j;\mu_j^\Lambda(\bm{\theta})  ,\sigma_j^\Lambda(\bm{\theta}))\times\exp(CS(\Lambda _j)),$$
  where $CS(\Lambda _j)$ denotes a cubic spline on $\Lambda_j$ with the aim to correct the approximation. 
  \item The third approach is the  simplified Laplace approximation and it is based on a series of expansions of ${\pi _{LA}}({\Lambda _j}|{\bm{\theta}},\bm{y},\bm{t},\bm{\delta} )$ around $\Lambda_j=\mu_j^\Lambda(\bm{\theta})$. This approach is very fast and the use of it can correct the location and scale of Gaussian approximation. 
\end{enumerate}
%\section{Dynamic prediction}

\section{Analysis of the HIV/AIDS data}
\subsection{Data description}
Three major electronic databases are maintained by the Brazilian National AIDS Program \cite{fonseca2010accuracy}: (i) SINAN-AIDS (Information System for Notifiable Diseases of AIDS Cases), the most important electronic AIDS surveillance database, which contains all reported cases since 1980; (ii) SISCEL (Laboratory Test Control System), designed to monitor laboratory tests such as CD4 counts and viral load tests for HIV/AIDS patients in the public health sector since 2002; and (iii) SICLOM (System for Logistic Control of Drugs) was developed to manage the logistics of AIDS treatment deliveries. Since 2002, it has been sharing the patient list with SISCEL. These three databases have previously been merged into a single database containing both HIV and AIDS cases using a process called record linkage, which was implemented by the Surveillance Unit of the Brazilian National AIDS Program \cite{fonseca2010accuracy}. This linkage strategy has been increasingly used in AIDS surveillance and research \cite{deapen2007population} to verify underreporting of cases and eliminate duplicates. In Brazil, this procedure has improved the quality of HIV/AIDS data information \cite{fonseca2010accuracy}. Notice that the period from 2002 to 2006 can be considered the first period with substantial information on both HIV/AIDS survival and CD4 exams. During this period, 88 laboratories across all 27 Brazilian states were utilizing SISCEL, encompassing 90\% of all CD4 and viral load tests conducted by the public health sector. Cases diagnosed before 2002 were excluded because personal identifiers were not available in the mortality database for the entire country before that date \cite{fonseca2010accuracy}. For institutional reasons, we only had access to a simple random sample of the combined database, which will be referred to as the HIV/AIDS data \cite{martins2016bayesian,martins2017joint}. The dataset consists of 500 random samples, denoted as N=500, from the original data. The explanatory variables in this study include age (1: older than 50, 0: younger than 50), gender (1: male, 0: female), PrevOI (1: with PrevOI, 0: without PrevOI), and time. \\
Figure \ref{surv} presents the Kaplan-Meier survival curve illustrating mortality over a five-year follow-up in HIV/AIDS data.
The data includes 34 deaths, accounting for 93.2\% of the right-censored cases. There are 298 (59.6\%) male patients, 60 (12\%) of whom are older than 50, and 198 (39.6\%) have a history of previous infection. The study includes 2757 longitudinal observations, with an average of 5.51 replications for each patient. The average time for censored patients is 997.9 days, whereas for the others, it is 862.4 days. On average, there are 18.52 patients with a standard deviation of 41.08 in each of the 27 Brazilian states. Additionally, two states have no patients in this dataset.

\begin{figure}[ht]
\begin{center}
\includegraphics[width=6cm]{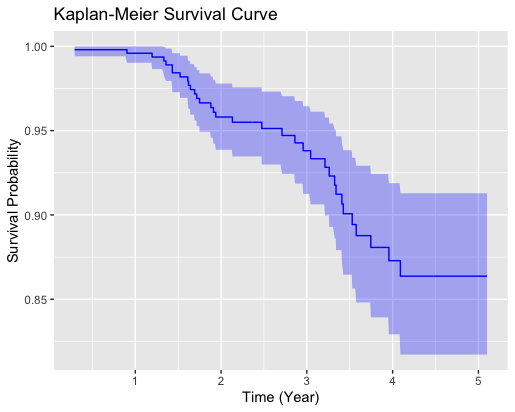}
\caption{Kaplan-Meier survival curve for mortality over five years of follow-up in HIV/AIDS data. }\label{surv}
\end{center}
\end{figure}
\FloatBarrier

\begin{figure}[ht]
\begin{center}
\includegraphics[width=6cm]{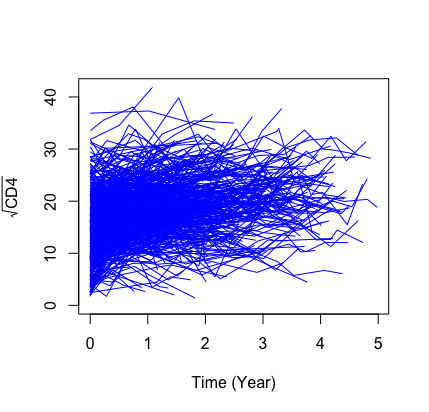}
\caption{Individual trajectories of the square root of CD4 levels over time in HIV/AIDS longitudinal data. }\label{spa}
\end{center}
\end{figure}
\FloatBarrier
\subsection{Data analysis}
In this section, the proposed methodology described in Section 3 is used to analysis the HIV/AIDS data.
As in \cite{martins2016bayesian} and \cite{martins2017joint}, the response variable is defined to be $\sqrt{CD4}$ in order to deal with the asymmetry in the longitudinal measurements. 
Figure \ref{spa} presents individual trajectories of square root-transformed CD4 counts over time in the HIV/AIDS dataset.
Let $Y_{ikj}$ denote the response variable for the $j^{th}$ time point of the $i^{th}$ patient in the $k^{th}$ state, $k=1,\cdots,K$. The following longitudinal model is considered to analyse the data:
\begin{eqnarray}\label{app1}
Y_{ikj}=\beta_0+\beta_1 \mbox{Age}_{ik}+\beta_2 \mbox{Gender}_{ik}+\beta_3 \mbox{PrevOI}_{ik}+g(s_{ikj})+\bm{z}_{ik}^\prime(s_{ikj}) \bm{b}_{ik} +\epsilon_{ikj},
  \end{eqnarray}
Also, for survival time, the following model is considered:
\begin{eqnarray}\label{app2}
h(t_{ik})=h_0(t_{ik}) \exp(\alpha_0+\alpha_1 \mbox{Age}_{ik}+\alpha_2 \mbox{Gender}_{ik}+\alpha_3 \mbox{PrevOI}_{ik}+\bm{\gamma}^\prime \bm{b}_{ik}+\nu_k)
  \end{eqnarray}
where $\epsilon_{ikj}\sim N(0,\sigma^2)$,
$(b_{ik0}, b_{ik1})^\prime \sim N_2(\bm{0},\bm{D})$, $g(s_{ikj})$ is a non-linear function of time, $h_0(t_{ik})=r t_{ik}^{r-1}$ and $\nu_k$ is distributed  as in \eqref{car10}. 
A variety of fitted joint models with different forms for the longitudinal association ($\bm{z}_{ik}^\prime(s_{ikj})\bm{b}_{ik}$), for the linkage structure  ($\bm{\gamma}^\prime \bm{b}_{ik}$)
and for the spatial random effects ($\nu_k$) are considered. Table  \ref{compar} reports the values of DIC and WAIC for purposes of the model comparison. Based on the results, model xi with 
$\bm{z}_{ik}^\prime(s_{ikj})=b_{ik0}+b_{ik1}s_{ikj}$,
$\bm{\gamma}^\prime \bm{b}_{ik}=\gamma_1 b_{ik0}+\gamma_2 b_{ik1}$
and by considering $\nu_k$ as the spatial random effects is the best fitting model. The results of this model can be found in Table \ref{app}. Dimensionality of the basis (i.e.: number of knots + degree) for the B-spline is considered to be equal to 5. 
The results in this table show that age, time and PrevOI are significant variables in the longitudinal model such that the older people or having 
PrevOI lead to smaller values of CD4 count measurements. Also, PrevOI is a significant variable in the survival model.  $\gamma_1$  and $\gamma_2$ are also significant which confirms the association between two outcomes.\\
In Figure \ref{fig2}, patient-specific predictions \cite{van2019new} for longitudinal trajectories and survival functions are depicted for three randomly selected patients. The left panel displays the observed values of $\sqrt{CD4}$ for each patient, represented by solid black lines, alongside their corresponding model-based predictions shown as blue dashed lines. The close proximity of these lines suggests high accuracy in the predictions. %This observation is further supported by panel (a) in Figure \ref{fig3}.\\
In the right panel of Figure \ref{fig2}, the survival curves superimposed by median lines for each patient are also given.\\
For model assessment and checking adequacy of the models, in addition of comparison between predicted values and observed values, a residual analysis is performed. For the residual analysis of the longitudinal component, the standardized marginal residuals are computed as 
\begin{eqnarray*}
e_{ikj}^{sm}=\frac{Y_{ikj}-\hat{\eta}^y_{ikj}}{\sqrt{\bm{z}_{ik}^\prime(s_{ikj})\hat{\bm{D} }\bm{z}_{ik}(s_{ikj})+\hat{\sigma^2}}},
 \end{eqnarray*}
where, $\hat{\eta}^y_{ikj}=\hat{\beta}_0+\hat{\beta}_1 \mbox{Age}_{ik}+\hat{\beta}_2 \mbox{Gender}_{ik}+\hat{\beta}_3 \mbox{PrevOI}_{ik}+\hat{g}(s_{ikj})$, such that $\hat{\beta}_k,~k=0,\cdots,3$ are estimated values of their parameters, $\hat{\bm{D} }$ and $\hat{\sigma^2}$ are estimated values of ${\bm{D} }$ and ${\sigma^2}$, respectively and $\hat{g}(s_{ikj})$ is the estimated spline function.  Panel (b) of Figure \ref{fig3} shows the standardized marginal residuals and supports the goodness of fit of the model. \\
For the residuals analysis of the spatial survival model, the Cox-Snell residuals are considered \cite{cox1968general}.
We have $r_{ik}^{CS}(t|\bm{\theta})=-\log S(t_{ik}|\bm{\theta})$ which can be calculated by the expected value over the posterior distribution as follows \cite{rizopoulos2011bayesian}:
$$r_{ik}^{CS}(t)=\int{r_{ik}^{CS}(t|\bm{\theta})\pi(\bm{\theta}|\bm{t},\bm{y})d\bm{\theta}},$$
In practice, $r_{ik}^{CS}(t)$ can be computed at the observed survival time $t_{ik}$. For checking the fit of the survival model, taking into account the censoring time,  the associated Kaplan–Meier estimate of the $r_{ik}^{CS}(t)$
is compared with
the survival function of the unit exponential distribution \cite{cox1968general}.          
The plot of the Cox-Snell residuals
is presented in Figure \ref{fig4}. The closeness of the two curves confirms the goodness of fit of the proposed survival model in the joint model.\\
Figure \ref{fig5} shows two maps for the HIV/AIDS data of  Brazilian states using model xi. The left panel of this figure represents the posterior spatial
mean risk, that is, if we define $\hat{\lambda}_{ik}=\hat{\alpha}_0+\hat{\alpha}_1 \mbox{Age}_{ik}+\hat{\alpha}_2 \mbox{Gender}_{ik}+\hat{\alpha}_3 \mbox{PrevOI}_{ik}+\hat{\bm{\gamma}}^\prime \hat{\bm{b}}_{ik}+\hat{\nu}_k$, then
 the posterior spatial
mean risk for each region is $\hat{\lambda}_{k}=\frac{\sum_{i=1}^{n_k}\hat{\lambda}_{ik}}{n_k},~k=1,\cdots,K=27$.
In this equation the hat sign for parameters denotes the approximate Bayesian estimate and the hat sign for random effects denotes the predicted values.  
The right panel of this figure
shows  the mean of the predicted values for $exp(\nu_k),~k=1,\cdots,K=27$.
The map of the relative risks shows that the states of the north of Brazil have higher level of HIV/AIDS risks. This conclusion was also obtained by \cite{martins2016bayesian}. \\
{Using a 12th Gen Intel Core i7-12700H processor and 32 GB of RAM, we assessed the computational efficiency of two methods—approximate Bayesian inference via \texttt{INLA} and Gibbs sampling via BUGS—specifically applied to Model xi. The INLA-based approximate Bayes approach completed the analysis in just 6.287 seconds, illustrating the advantages of a streamlined, highly efficient method well-suited to modern, high-performance hardware. By comparison, Gibbs sampling with 50,000 iterations and two chains in \texttt{R2OpenBUGS} required a considerably longer processing time of 412 minutes (roughly 6.87 hours) to achieve similar results. This striking contrast highlights the computational intensity of Gibbs sampling, particularly in probabilistic frameworks like \texttt{R2OpenBUGS}, where iterative sampling demands extensive processing time. Given these results, we omit further Gibbs sampling outputs to emphasize the efficiency gains achieved with the approximate Bayes approach. The BUGS code for this application can be found  in \url{https://github.com/tbaghfalaki/ASJM}. }

\begin{table}[hbt!]
 \centering
  \caption{\label{compar} Model comparison criteria for the candidate joint models. }
  \begin{tabular}{c|c|c|c|cc}
     \hline
     Model  &\ $\bm{z}_{ik}^\prime(s_{ikj}) \bm{b}_{ik}$     &\   $\bm{\gamma}^\prime \bm{b}_{ik}$     &\  $\nu_k$    &\   WAIC    &\  DIC     \\\hline
N &\ 0  &\ 0  &\ 0  &\ 17642 &\  17641 \\
i &\ $b_{ik0}$  &\ 0  &\ 0  &\ 15193 &\ 15171   \\
ii &\ $b_{ik0}$ &\ $\gamma_1 b_{ik0}$  &\ 0  &\ 17648 &\ 17643   \\
iii  &\ $b_{ik0}$  &\ $\gamma_1 b_{ik0}$ &\ $\nu_k$  &\ 15163 &\  15139  \\
iv   &\ 0  &\ 0  &\ $\nu_k$  &\ 17641 &\  17640  \\
v    &\ $b_{ik0}+b_{ik1}s_{ikj}$  &\ $\gamma_1 b_{ik0}$  &\ $\nu_k$  &\ 14069 &\  14008 \\
vi     &\ $b_{ik0}+b_{ik1}s_{ikj}$  &\ $\gamma_2 b_{ik1}$  &\ $\nu_k$  &\ 14047 &\  13989  \\
vii     &\ $b_{ik0}+b_{ik1}s_{ikj}$  &\ $\gamma_1 b_{ik0}$  &\ 0  &\ 14077  &\  14014 \\
viii    &\ $b_{ik0}+b_{ik1}s_{ikj}$  &\ $\gamma_2 b_{ik1}$  &\ 0  &\ 14036  &\  13977  \\
ix &\ $b_{ik0}+b_{ik1}s_{ikj}$  &\ $\gamma_1 b_{ik0}+\gamma_2 b_{ik1}$  &\ 0  &\ 14033 &\ 13976   \\
x  &\ $b_{ik0}+b_{ik1}s_{ikj}$  &\ 0  &\ $\nu_k$  &\ 14085 &\  14015  \\
xi &\ $b_{ik0}+b_{ik1}s_{ikj}$  &\ $\gamma_1 b_{ik0}+\gamma_2 b_{ik1}$ &\ $\nu_k$  &\ 14025 &\ 13967  \\
      \hline
     \end{tabular}
\end{table}

\begin{table}[hbt!]
 \centering
 \footnotesize
  \caption{\label{app} Bayesian parameter estimates, posterior means (standard deviations), and $95\%$ credible intervals for
the  HIV/AIDS data using model xi.}
  \begin{tabular}{c|cc|cc|c}
     \hline
   \multirow{2}{*}{Parameter } &\  \multirow{2}{*}{Mean} &\ \multirow{2}{*}{SD} & \multicolumn{2}{c|}{Credible Interval}  &\  \multirow{2}{*}{Median}  \\\cline{4-5}
        &\          &\  &\ $2.5\%$ &\ $97.5\%$  &\    \\\hline
    $\beta_0$ &\ 9.443 &\ 0.899 &\ 7.68 &\ 11.205 &\ 9.443  \\
Age ($\beta_1$) &\ -1.442 &\ 0.675 &\ -2.766 &\ -0.118 &\ -1.442  \\
Gender ($\beta_2$) &\ -0.821 &\ 0.45 &\ -1.704 &\ 0.062 &\ -0.821  \\
PrevOI ($\beta_3$) &\ -1.568 &\ 0.456 &\ -2.462 &\ -0.675 &\ -1.568  \\
g(s) &\ -34.46 &\ 3.989 &\ -42.283 &\ -26.636 &\ -34.459  \\
  &\ 11.123 &\ 2.414 &\ 6.388 &\ 15.857 &\ 11.123  \\
  &\ -46.778 &\ 4.909 &\ -56.403 &\ -37.149 &\ -46.779  \\
$\sigma^{-2}$ &\ 0.159 &\ 0.006 &\ 0.148 &\ 0.17 &\ 0.159  \\
$\alpha_0$ &\ -5.138 &\ 0.407 &\ -5.992 &\ -4.394 &\ -5.118  \\
Age ($\alpha_1$) &\ 0.835 &\ 0.428 &\ -0.046 &\ 1.636 &\ 0.849  \\
Gender ($\alpha_2$) &\ 0.352 &\ 0.377 &\ -0.364 &\ 1.117 &\ 0.343  \\
PrevOI ($\alpha_3$) &\ 0.89 &\ 0.366 &\ 0.185 &\ 1.622 &\ 0.885  \\
$r$ &\ 1.159 &\ 0.027 &\ 1.108 &\ 1.214 &\ 1.159  \\
$\tau^{-1}$  &\ 193.241 &\ 35.064 &\ 127.947 &\ 264.402 &\ 192.497  \\
 $\gamma_1$  &\ -0.138 &\ 0.04 &\ -0.218 &\ -0.061 &\ -0.137  \\
 $\gamma_2$  &\ -0.376 &\ 0.091 &\ -0.555 &\ -0.196 &\ -0.377  \\
$D_{11}^{-1}$ &\ 0.039 &\ 0.003 &\ 0.033 &\ 0.045 &\ 0.039  \\
$D_{22}^{-1}$ &\ 0.168 &\ 0.018 &\ 0.136 &\ 0.207 &\ 0.167  \\
$\rho$ &\ -0.354 &\ 0.056 &\ -0.457 &\ -0.238 &\ -0.356  \\       \hline
     \end{tabular}
\end{table}

    \begin{figure}[hbt!]
\centering
\includegraphics[width=12cm]{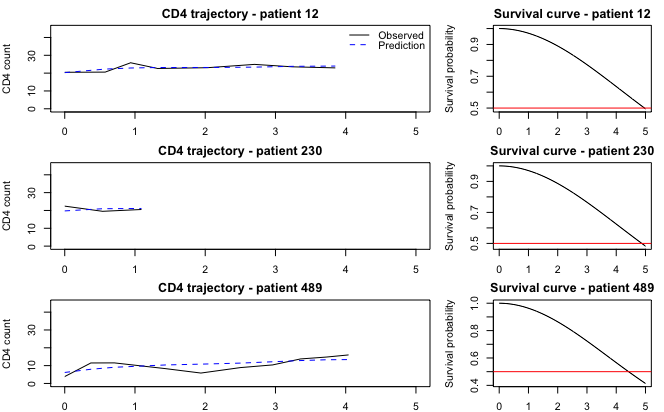}
\vspace*{-.1cm} \caption{\label{fig2} Plots of observed and predicted values for longitudinal outcomes (panel a) and standardized marginal residuals (panel b). }
\end{figure}

\begin{figure}[hbt!]
\centering
\includegraphics[width=9cm]{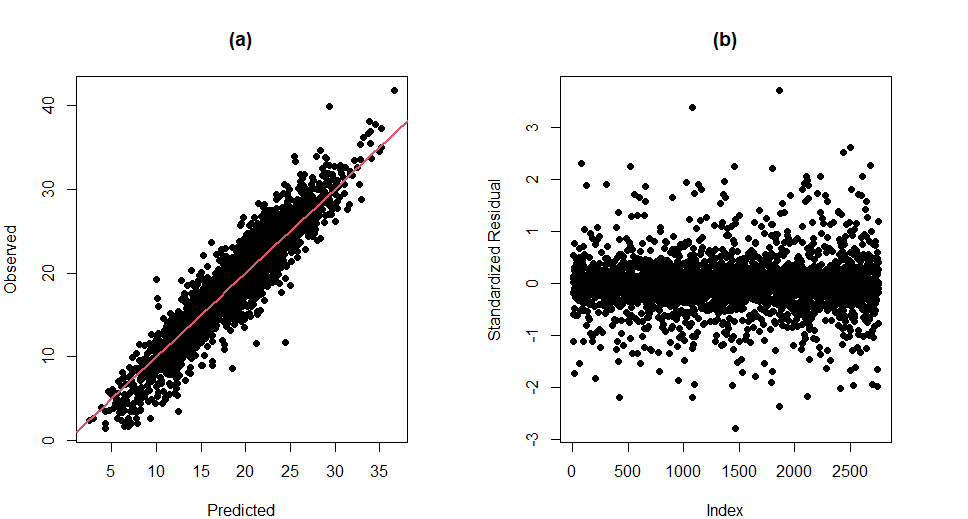}
\vspace*{-.1cm} \caption{\label{fig3} Plots of observed and predicted values for longitudinal outcomes (panel a) and standardized marginal residuals (panel b). }
\end{figure}

\begin{figure}[hbt!]
\centering
\includegraphics[width=6cm]{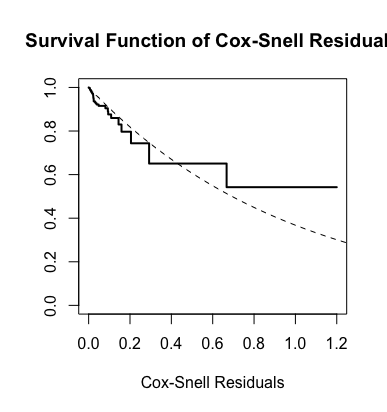}
\vspace*{-.1cm} \caption{\label{fig4}  
The empirical survival curves based on the  Kaplan-Meier posterior estimates of Cox-Snell residuals (solid line) and the unit exponential distribution (dashed line).  }
\end{figure}

\begin{figure}[hbt!]
\centering
\includegraphics[width=9cm]{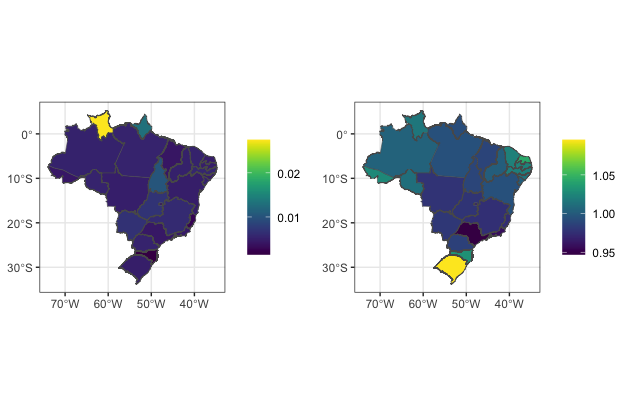}
\vspace*{-.1cm} \caption{\label{fig5}  
Map of the spatial mean (left panel) and relative risks (right panel) using model xi.} 
\end{figure}
\FloatBarrier

\section{Simulation studies}
\subsection{Scenario 1}
In this section, we conducted simulation studies to assess the performance of the proposed joint model. When considering an areal data set, the shapefile of North Carolina from the \texttt{spData}, \texttt{maptools}, and \texttt{spdep} packages in R is utilized. There are $K=100$ regions in this map (see figure \ref{fig1}).\\
We consider two different levels of sample sizes including $n_k=20$ and $50$, $k=1,\cdots,K$. 
%By considering this structure, the total number of sample sizes are $n=1000$ and $2000$, respectively. 
We consider two explanatory variables for generating the data: $\bm{x}_1$ is generated from the standard normal distribution as continuous random variable and $\bm{x}_2$ is generated from the Bernoulli distribution with a success probability of $0.5$. \\
The following linear mixed effects model is considered for the longitudinal model:
 \begin{eqnarray}
{Y_{ikj}} = {\beta _0} + {\beta _1}{x_{1ik}} + {\beta _2}{x_{2ik}} + g({t_{ikj}}) + {b_{0ik}} + {b_{1ik}}{t_{ikj}} + {\varepsilon _{ikj}},
 \end{eqnarray}
where $t_{ik}=(0,0.05,0.1,\cdots,1)$, $\bm{\beta}=(\beta _0,\beta _1,\beta _2)^\prime=(2,-1,1)^\prime$, $g(t)=sin(2\pi t)$, $\varepsilon _{ikj}\sim N(0,\sigma^2)$, $\sigma^2=1$, 
$\bm{b}_{ik}=({b_{0ik}},{b_{1ik}})^\prime \sim {N_2}(\bm{0},{\bm{D}^{ - 1}})$, where  $\bm{D}^{-1}= \left( {\begin{array}{*{20}{c}}
    {D_{11}^{-1}} & {\frac{\rho}{\sqrt{ {D_{11}^{-1}} {D_{22}^{-1}}}}}  \\
   {\frac{\rho}{\sqrt{ {D_{11}^{-1}} {D_{22}^{-1}}}}} &  {D_{22}^{-1}} \\
\end{array}} \right)$ is the precision matrix such that $\bm{D} = \left( {\begin{array}{*{20}{c}}
   1 & {0.5}  \\
   {0.5} & 1  \\
\end{array}} \right)$. 
For the survival time the following model is considered:
 \begin{eqnarray}
h_{ik}(t)=h_{0ik}(t)\exp(\alpha_0+\alpha_1 {x_{1ik}}+\gamma_1{b_{0ik}} + \gamma_2{b_{1ik}}+\nu_k),
 \end{eqnarray}
where $\bm{\alpha}=(\alpha_0,\alpha_1)^\prime=(0.5,-0.5)^\prime$, $\gamma_1=1$, $\gamma_2=-1$, $h_{0ik}(t)=\varpi t^{\varpi-1}$, $\varpi=1$ and $40\%$ rate of right censoring is considered. For generating $\bm{\nu}=(\nu_1,\cdots,\nu_K)^\prime$, equation \eqref{car10} %e2} 
 is considered such that $\tau=0.1$\\
The following four models are assessed with  $M=200$ replications:
\begin{description}
    \item[Model I] Separate models without spatial random effects. 
    \item[Model II] Separate models with spatial random effects. 
    \item[Model III] Joint model without spatial random effects. 
    \item[Model IV] Joint model with spatial random effects. 
\end{description}
The  first two models ignore the joint analysis of two outcomes. The first and the third ones ignore the spatial
effects and model IV is the model by considering spatial effects and joint analysis. For model comparison  relative bias and root of mean squared errors are used such that
$ Rbias\left( \vartheta  \right) = \frac{{{\bar{\hat{ \vartheta}}}}}{\vartheta } - 1$, $RMSE(\vartheta ) = \sqrt {\frac{{\sum\limits_{l = 1}^M {{{({{\hat \vartheta }_l} - \vartheta )}^2}} }}{M}}$, where $ {{\hat \vartheta }_l} $ is the estimated value of parameter $ \vartheta $ for the $l$th simulation run and  $ \bar{\hat{ \vartheta}} = \dfrac{{\sum\limits_{l = 1}^M {{{{{\hat \vartheta }_l}}}}}}{M}$. Also, 
DIC and WAIC are utilized for selecting the best-fitting model. {A difference in DIC greater than 10 is regarded as a strong indicator of model preference, whereas a difference between 5 and 10 suggests meaningful distinctions, indicating that the models may still be comparable \cite{spiegelhalter2002bayesian}. Similarly, for WAIC, a difference of 10 is considered substantial evidence in favor of the model with the lower WAIC \cite{gelman2014understanding}.}
\\
The results shown in Table \ref{sim0} and Figure \ref{sen1} indicate that the four models exhibit significant differences in their performance for parameter estimation. Model I (separate models without spatial effects) exhibits the highest bias and RMSE, resulting in unreliable estimates. Model II (separate models with spatial effects) shows some improvement, but its accuracy remains insufficient. In contrast, Model III (joint model without spatial effects) outperforms the separate models; however, neglecting spatial effects can still lead to inaccurate estimates. Model IV (joint model with spatial effects) delivers the best performance by effectively incorporating spatial correlations and jointly analyzing both outcomes. The DIC and WAIC results further reinforce the superiority of Model IV, with substantial differences exceeding 10 in these metrics, indicating its excellent fit to the data. Overall, Model IV emerges as the optimal choice for the analyzed data, highlighting the critical importance of considering spatial effects and joint analysis for achieving more accurate parameter estimates.\\
In terms of variance components, the estimates for $\sigma^2$, $D_{11}^{-1}$, $D_{22}^{-1}$, and $\rho$ across all models are close to the true values, with small Rbias and RMSE values reflecting good accuracy. The estimated values of spatial precision ($\tau^{-1}$) for Models II and IV are also close to the true values, with Model IV yielding the closest estimate. The parameters $\gamma_1$ and $\gamma_2$ in Model IV are near the true values, whereas Model III shows considerable deviation with larger Rbias and RMSE values. Based on the DIC and WAIC metrics, Model IV is confirmed as the best-fitting model.\\
{A comparison of parameter estimates across different sample sizes indicates that the model with $n_k = 50$ generally outperforms the one with $n_k = 20$. For instance, the relative bias and RMSE for the parameter $\beta_0$ are lower with $n_k = 50$ (Rbias = 0.020, RMSE = 0.042) compared to $n_k = 20$ (Rbias = 0.021, RMSE = 0.047), suggesting improved accuracy in estimation. Although the relative bias for $\beta_1$ is slightly higher at $n_k = 50$ (Rbias = 0.005), its RMSE is lower (0.015) compared to $n_k = 20$ (0.021), indicating better overall precision. The parameter $\alpha_0$ demonstrates a less negative relative bias with $n_k = 50$ (-1.199) and a comparable RMSE (0.600), reflecting greater stability. Notably, the RMSE for the variance parameter $\sigma^2$ shows a significant reduction from $n_k = 20$ (0.009) to $n_k = 50$ (0.006), indicating enhanced estimation accuracy. Furthermore, the relative bias for $D_{11}^{-1}$ is virtually eliminated at $n_k = 50$ (Rbias = 0.000), with a marked improvement in RMSE from 0.041 to 0.024. Collectively, these findings underscore the importance of larger sample sizes in obtaining more reliable parameter estimates.}\\
This simulation study was conducted on a MacBook Pro 2020, equipped with Apple's state-of-the-art M1 chip and 256GB of storage. In Table \ref{sim0}, we provide the mean and standard deviation of the computational times measured in minutes. Notably, despite the complexity of the models, the computational times remain relatively short. For more intricate models, although the computational time increases, it typically does not exceed one minute on average for $n_k=50$.\\
Figure \ref{n110}  shows  plots of the estimated curves superimposed by its $95\% CI$ for the non-linear function of time  for $n_k=20$ (panel a) and for $n_k=50$ (panel b). These figures  confirm our discussions about  the strategy about the use of a spline function  for take into account the non-linear effect of time, i.e., the estimated curves for the  B-spline are near to the real curves and by increasing the $n_k$ the length of their $95\% CI$  decreases.

\begin{figure}[hbt!]
\centering
\includegraphics[width=12cm]{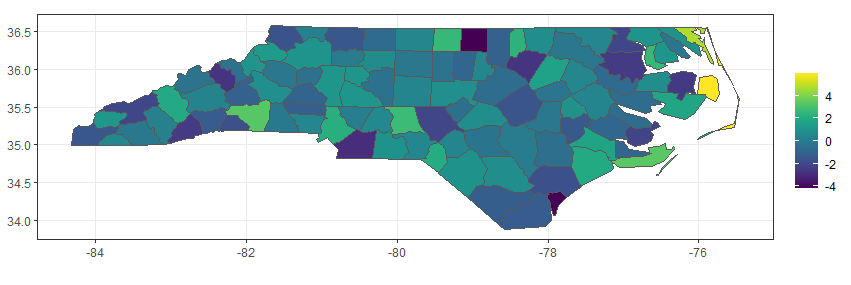}
\vspace*{-.1cm} \caption{\label{fig1}  
Generated spatial frailties of North Carolina for one replication  in the simulation study.
 }
\end{figure}

\begin{figure}[hbt!]
\centering
\includegraphics[width=12cm]{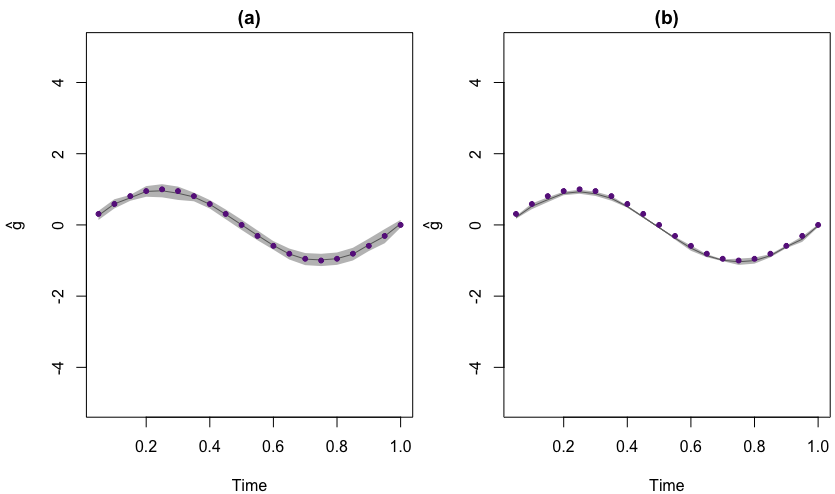}
\vspace*{-.1cm} \caption{\label{n110}
Plot of the estimated curve superimposed by its $95\%$ CI for the B-spline for different values of $n_k,~k=1,\cdots,K=100$. (a): $n_k=20$, (b):  $n_k=50$.   }
\end{figure}
\FloatBarrier

\begin{figure}[hbt!]
\centering
\includegraphics[width=15cm]{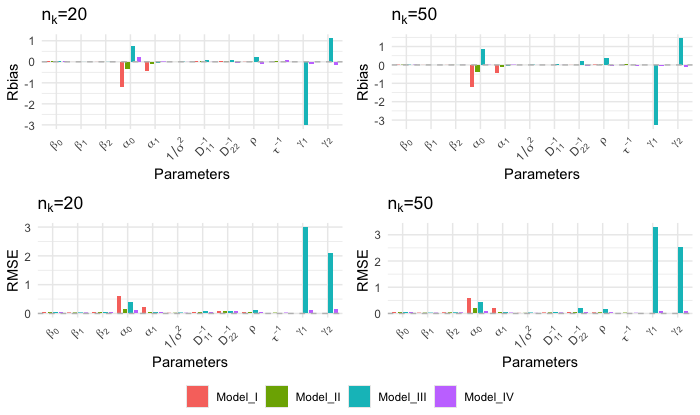}
\vspace*{-.1cm} \caption{\label{sen1}  
Comparison of Rbias and RMSE  across four models (model I to model IV) in scenario 1 of the simulation study. 
 }
\end{figure}

%\begin{landscape}
\begin{sidewaystable}
 \centering
 \tiny
  \caption{\label{sim0} Simulation Study Results for the Spatial Joint Model: Parameter Estimates, Standard Errors, Relative Bias, and RMSE for Scenario 1 with M=100 Simulations and $n_k=20,~50$. }
  \begin{tabular}{c|c|c|cccc|cccc|cccc|cccc}
     \hline
           &     &   & \multicolumn{4}{c}{Model I}   & \multicolumn{4}{|c}{Model II}    & \multicolumn{4}{|c}{Model III}   & \multicolumn{4}{|c}{Model IV} \\\hline
$n_k$ &\ Parameter &\ Real  &\ Est.  &\ S.E.  &\ Rbias &\ RMSE &\ Est.  &\ S.E.  &\ Rbias &\ RMSE &\ Est.  &\ S.E.  &\ Rbias &\ RMSE  &\ Est.  &\ S.E.  &\ Rbias &\ RMSE\\\hline
   & $\beta_0$ & 2.000 & 2.041 & 0.041 & 0.021 & 0.047 & 2.041 & 0.041 & 0.021 & 0.047 & 2.048 & 0.040 & 0.024 & 0.051 & 2.042 & 0.040 & 0.021 & 0.047 \\ 
   & $\beta_1$  & -1.000 & -1.001 & 0.027 & 0.001 & 0.021 & -1.001 & 0.027 & 0.001 & 0.021 & -0.997 & 0.026 & -0.003 & 0.021 & -0.999 & 0.027 & -0.001 & 0.021 \\ 
   & $\beta_2$  & 1.000 & 0.996 & 0.056 & -0.004 & 0.042 & 0.996 & 0.056 & -0.004 & 0.042 & 0.993 & 0.054 & -0.007 & 0.043 & 0.997 & 0.054 & -0.003 & 0.042 \\ 
& $\alpha_0$ & 0.500 & -0.103 & 0.090 & -1.206 & 0.603 & 0.336 & 0.054 & -0.328 & 0.164 & 0.887 & 0.062 & 0.774 & 0.387 & 0.607 & 0.049 & 0.214 & 0.117 \\ 
& $\alpha_1$ & -0.500 & -0.290 & 0.039 & -0.420 & 0.210 & -0.443 & 0.040 & -0.115 & 0.058 & -0.461 & 0.049 & -0.078 & 0.051 & -0.486 & 0.038 & -0.028 & 0.034 \\ 
& $1/\sigma^{2}$ & 1.000 & 0.999 & 0.012 & -0.001 & 0.009 & 0.999 & 0.012 & -0.001 & 0.009 & 0.981 & 0.011 & -0.019 & 0.019 & 1.004 & 0.012 & 0.004 & 0.010 \\ 
20  & $D_{11}^{-1}$ & 1.000 & 1.015 & 0.048 & 0.015 & 0.041 & 1.016 & 0.048 & 0.016 & 0.041 & 1.073 & 0.050 & 0.073 & 0.075 & 1.007 & 0.046 & 0.007 & 0.038 \\ 
  & $D_{22}^{-1}$  & 1.000 & 1.014 & 0.087 & 0.014 & 0.071 & 1.015 & 0.088 & 0.015 & 0.071 & 1.098 & 0.072 & 0.098 & 0.102 & 0.938 & 0.064 & -0.062 & 0.071 \\ 
& $\rho$ & 0.500 & 0.495 & 0.057 & -0.011 & 0.043 & 0.494 & 0.057 & -0.011 & 0.042 & 0.610 & 0.027 & 0.220 & 0.110 & 0.448 & 0.042 & -0.104 & 0.056 \\ 
& $\tau^{-1}$ & 0.100 & - & - & -  & - & 0.105 & 0.018 & 0.049 & 0.014 &  - & - & -  & - & 0.093 & 0.015 & -0.070 & 0.017 \\ 
& $\gamma_1$  & 1.000 &  - & - & -  & - &  - & - & - & - & 1.998 & 0.168 & -2.998 & 2.998 & 0.887 & 0.059 & -0.113 & 0.113 \\ 
& $\gamma_2$  & -1.000 &  - & - & -  & - &  - & - & -  & - & -2.109 & 0.174 & 1.109 & 2.109 & -0.834 & 0.056 & -0.166 & 0.166 \\ 
& DIC   &   & \multicolumn{4}{c}{54138}   & \multicolumn{4}{|c}{51593}    & \multicolumn{4}{|c}{52383}   & \multicolumn{4}{|c}{50930} \\ 
& WAIC    &   & \multicolumn{4}{c}{54228}   & \multicolumn{4}{|c}{51778}    & \multicolumn{4}{|c}{54320}   & \multicolumn{4}{|c}{51258} \\ 
& Computational time    &    & \multicolumn{4}{c}{0.100 (0.212)}   & \multicolumn{4}{|c}{0.134 (0.042)}    & \multicolumn{4}{|c}{0.205 (0.052)}   & \multicolumn{4}{|c}{0.448 (0.476)} \\\hline
   & $\beta_0$  & 2.000 & 2.041 & 0.026 & 0.020 & 0.042 & 2.041 & 0.026 & 0.020 & 0.042 & 2.041 & 0.025 & 0.020 & 0.043 & 2.040 & 0.025 & 0.020 & 0.042 \\ 
   & $\beta_1$  & -1.000 & -1.005 & 0.018 & 0.005 & 0.015 & -1.005 & 0.018 & 0.005 & 0.015 & -1.001 & 0.018 & 0.001 & 0.015 & -1.002 & 0.018 & 0.002 & 0.015 \\ 
   & $\beta_2$  & 1.000 & 0.995 & 0.036 & -0.005 & 0.029 & 0.995 & 0.036 & -0.005 & 0.029 & 0.997 & 0.036 & -0.003 & 0.029 & 0.996 & 0.036 & -0.004 & 0.028 \\ 
& $\alpha_0$ & 0.500 & -0.100 & 0.083 & -1.199 & 0.600 & 0.308 & 0.038 & -0.383 & 0.192 & 0.923 & 0.056 & 0.846 & 0.423 & 0.507 & 0.040 & 0.014 & 0.097 \\ 
& $\alpha_1$ & -0.500 & -0.290 & 0.027 & -0.420 & 0.210 & -0.438 & 0.021 & -0.125 & 0.062 & -0.466 & 0.032 & -0.068 & 0.038 & -0.489 & 0.019 & -0.022 & 0.018 \\ 
& $1/\sigma^2$ & 1.000 & 1.001 & 0.007 & 0.001 & 0.006 & 1.001 & 0.008 & 0.001 & 0.006 & 0.980 & 0.007 & -0.020 & 0.020 & 1.002 & 0.007 & 0.002 & 0.006 \\ 
50  & $D_{11}^{-1}$ & 1.000 & 0.999 & 0.030 & -0.001 & 0.024 & 1.000 & 0.032 & 0.000 & 0.026 & 1.073 & 0.040 & 0.073 & 0.073 & 0.996 & 0.029 & -0.004 & 0.024 \\ 
  & $D_{22}^{-1}$  & 1.000 & 0.993 & 0.054 & -0.007 & 0.046 & 0.995 & 0.053 & -0.005 & 0.045 & 1.195 & 0.085 & 0.195 & 0.196 & 0.964 & 0.046 & -0.036 & 0.047 \\ 
& $\rho$ & 0.500 & 0.486 & 0.033 & -0.028 & 0.028 & 0.489 & 0.036 & -0.023 & 0.030 & 0.686 & 0.027 & 0.372 & 0.186 & 0.471 & 0.027 & -0.058 & 0.034 \\ 
& $\tau^{-1}$ & 0.100 &  - & - & -  & - & 0.108 & 0.020 & 0.075 & 0.016 &  - & - & -  & - & 0.095 & 0.016 & -0.050 & 0.011 \\ 
& $\gamma_1$  & 1.000 &  - & - & -  & - &  - & - & -  & - &  2.258 & 0.553 & -3.258 & 3.292 & 0.919 & 0.036 & -0.081 & 0.082 \\ 
& $\gamma_2$  & -1.000 &  - & - & -  & - &  - & - & -  & - & -2.438 & 0.699 & 1.438 & 2.520 & -0.891 & 0.041 & -0.109 & 0.109 \\ 
& DIC   &   & \multicolumn{4}{c}{135283}   & \multicolumn{4}{|c}{128923}    & \multicolumn{4}{|c}{130889}   & \multicolumn{4}{|c}{127146} \\ 
& WAIC    &   & \multicolumn{4}{c}{134404}   & \multicolumn{4}{|c}{129249}    & \multicolumn{4}{|c}{136254}   & \multicolumn{4}{|c}{127908} \\ 
& Computational time    &    & \multicolumn{4}{c}{0.250 (0.052)}   & \multicolumn{4}{|c}{0.425 (0.812)}    & \multicolumn{4}{|c}{0.640 (0.458)}   & \multicolumn{4}{|c}{0.803 (0.437)} \\\hline
     \end{tabular}
\end{sidewaystable}
%\end{landscape}
\FloatBarrier

{
\subsection{Scenario 2}
In this scenario, we aim to replicate the data using real values similar to those obtained in the application section. For this purpose, we will utilize models \eqref{app1} and \eqref{app2}, where Age is generated from a standard normal distribution, and both Gender and PrevOI are generated from a Bernoulli distribution with a success probability of 0.5. The same function as in the previous simulation study, \( g(t) = \sin(2\pi t) \), will be used.\\
The parameters are defined as follows:
$\bm{\beta} = (\beta_0, \beta_1, \beta_2, \beta_3)^\prime = (9, -1, -1, -1.5)^\prime,$
$\bm{\alpha} = (\alpha_0, \alpha_1, \alpha_2, \alpha_3)^\prime = (-6, 0.5, 0.5, 1)^\prime,$
$\gamma_1 = -0.2, \quad \gamma_2 = -0.5$, and $\varpi = 2$.
The random effects are defined as \( \bm{b}_{ik} = (b_{0ik}, b_{1ik})^\prime \sim N_2(\bm{0}, \bm{D}^{-1}) \), where
\[
\bm{D} = \begin{pmatrix}
25 & -4 \\
-4 & 6
\end{pmatrix}
\]
resulting in \( D_{11}^{-1} = 0.04 \), \( D_{22}^{-1} = 0.167 \), \( \rho = -0.326 \), \( \sigma^2 = 5 \), and \( \tau = 12 \).\\
The results summarized in Table \ref{sim1} and Figure \ref{sen2}  present the performance of four models in estimating parameters within a spatial joint model framework under Scenario 2, using \(M=100\) simulations and a sample size of \(n_k=50\). Each model's estimates, standard errors (S.E.), relative bias (Rbias), and root mean squared error (RMSE) are reported for various parameters.\\
Model I (separate models without spatial effects) generally shows high relative biases and RMSEs for most parameters, particularly for \(\alpha_0\) and \(\varpi\), indicating less reliable estimates. For example, \(\alpha_0\) is estimated at \(-3.584\) with a significant negative Rbias of \(-0.403\) and an RMSE of \(2.416\). Model II (separate models with spatial effects) improves upon Model I, with more accurate estimates, particularly for the spatial variance parameters \(D_{11}^{-1}\) and \(D_{22}^{-1}\), which are estimated closer to their true values, although biases remain notable.
Model III (joint model without spatial effects) exhibits better parameter estimates than Models I and II, with smaller Rbias and RMSE values across several parameters, including \(\beta_0\) and \(\beta_2\). However, it still fails to fully leverage spatial correlations. Model IV (joint model with spatial effects) stands out as the most effective, yielding the closest estimates to the true values for most parameters, including \(\beta_0\) (est. \(9.045\), Rbias \(0.005\), RMSE \(0.163\)) and \(\alpha_1\) (est. \(0.478\), Rbias \(-0.044\), RMSE \(0.089\)).\\
The DIC and WAIC values further confirm the superiority of Model IV, with DIC at \(68136\) and WAIC at \(68281\), both significantly lower than those of the other models. This suggests that Model IV provides a substantially better fit to the data.\\
Regarding computational efficiency, Model I has the shortest average computational time (mean \(0.084\) minutes), while Model IV takes longer (mean \(0.422\) minutes). Despite this increase, the time remains relatively low, indicating that the complexity of the model does not excessively hinder performance.\\
Overall, the findings suggest that incorporating spatial effects and employing a  joint modeling approach significantly enhances parameter estimation accuracy. Model IV emerges as the best model, emphasizing the critical role of spatial correlations in analysis.

\begin{figure}[hbt!]
\centering
\includegraphics[width=9cm]{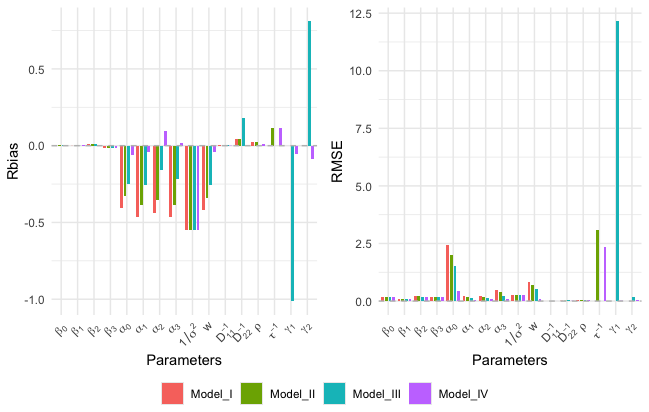}
\vspace*{-.1cm} \caption{\label{sen2}  
Comparison of Rbias and RMSE  across four models (model I to model IV) in scenario 1 of the simulation study. 
 }
\end{figure}
\FloatBarrier

%\begin{landscape}
\begin{sidewaystable}
 \centering
 \tiny
  \caption{\label{sim1} Simulation Study Results for the Spatial Joint Model: Parameter Estimates, Standard Errors, Relative Bias, and RMSE for Scenario 2 with M=100 Simulations and $n_k=50$. The computational time is measured in minutes and includes the mean (standard deviation).
  }
  \begin{tabular}{c|c|cccc|cccc|cccc|cccc}
     \hline
                &   & \multicolumn{4}{c}{Model I}   & \multicolumn{4}{|c}{Model II}    & \multicolumn{4}{|c}{Model III}   & \multicolumn{4}{|c}{Model IV} \\\hline
 Parameter &\ Real  &\ Est.  &\ S.E.  &\ Rbias &\ RMSE &\ Est.  &\ S.E.  &\ Rbias &\ RMSE &\ Est.  &\ S.E.  &\ Rbias &\ RMSE  &\ Est.  &\ S.E.  &\ Rbias &\ RMSE\\\hline
$\beta_0$ & 9.000 & 9.029 & 0.215 & 0.003 & 0.159 & 9.029 & 0.215 & 0.003 & 0.159 & 9.033 & 0.217 & 0.004 & 0.161 & 9.045 & 0.217 & 0.005 & 0.163 \\ 
  $\beta_1$ & -1.000 & -0.999 & 0.129 & -0.001 & 0.103 & -0.999 & 0.129 & -0.001 & 0.103 & -1.001 & 0.130 & 0.001 & 0.103 & -1.003 & 0.130 & 0.003 & 0.103 \\ 
$\beta_2$ & -1.000 & -1.012 & 0.246 & 0.012 & 0.198 & -1.012 & 0.246 & 0.012 & 0.198 & -1.009 & 0.246 & 0.009 & 0.197 & -1.006 & 0.246 & 0.006 & 0.197 \\ 
$\beta_3$ & -1.500 & -1.478 & 0.239 & -0.015 & 0.192 & -1.478 & 0.239 & -0.015 & 0.192 & -1.478 & 0.240 & -0.015 & 0.192 & -1.474 & 0.239 & -0.017 & 0.192 \\ 
$\alpha_0$ & -6.000 & -3.584 & 0.380 & -0.403 & 2.416 & -4.021 & 0.149 & -0.330 & 1.979 & -4.485 & 0.762 & -0.252 & 1.515 & -5.627 & 0.332 & -0.062 & 0.416 \\ 
$\alpha_1$ & 0.500 & 0.267 & 0.046 & -0.466 & 0.233 & 0.306 & 0.036 & -0.387 & 0.194 & 0.372 & 0.082 & -0.257 & 0.134 & 0.478 & 0.054 & -0.044 & 0.044 \\ 
$\alpha_2$ & 0.500 & 0.281 & 0.073 & -0.439 & 0.219 & 0.325 & 0.058 & -0.351 & 0.175 & 0.422 & 0.121 & -0.156 & 0.121 & 0.548 & 0.103 & 0.097 & 0.089 \\ 
$\alpha_3$ & 1.000 & 0.535 & 0.075 & -0.465 & 0.465 & 0.616 & 0.062 & -0.384 & 0.384 & 0.781 & 0.155 & -0.219 & 0.231 & 1.015 & 0.109 & 0.015 & 0.083 \\ 
$1/\sigma^{2}$ & 0.447 & 0.200 & 0.003 & -0.552 & 0.247 & 0.200 & 0.003 & -0.552 & 0.247 & 0.201 & 0.003 & -0.551 & 0.247 & 0.201 & 0.003 & -0.550 & 0.246 \\ 
 $\varpi$ & 2.000 & 1.156 & 0.127 & -0.422 & 0.844 & 1.321 & 0.045 & -0.340 & 0.679 & 1.487 & 0.262 & -0.257 & 0.513 & 1.920 & 0.101 & -0.040 & 0.099 \\ 
$D_{11}^{-1}$ & 0.040 & 0.040 & 0.002 & 0.002 & 0.001 & 0.040 & 0.002 & 0.001 & 0.001 & 0.040 & 0.002 & 0.002 & 0.001 & 0.040 & 0.002 & -0.002 & 0.001 \\ 
$D_{22}^{-1}$   & 0.167 & 0.174 & 0.015 & 0.045 & 0.012 & 0.174 & 0.015 & 0.045 & 0.012 & 0.196 & 0.226 & 0.179 & 0.039 & 0.166 & 0.015 & -0.004 & 0.011 \\ 
 $\rho$  & -0.327 & -0.334 & 0.037 & 0.023 & 0.031 & -0.334 & 0.037 & 0.022 & 0.032 & -0.326 & 0.078 & -0.003 & 0.043 & -0.330 & 0.039 & 0.010 & 0.033 \\ 
$\tau^{-1}$ & 12.000 &  - & - & -  & - & 13.414 & 3.157 & 0.118 & 3.087 & - & - & -  & - & 13.399 & 2.532 & 0.116 & 2.336 \\ 
$\gamma_1$ & -0.200 &  - & - & -  & - &  - & - & -  & - & -0.147 & 0.025 & -1.012 & 12.147 & -0.189 & 0.015 & -0.054 & 0.015 \\ 
  $\gamma_2$ & -0.500 &  - & - & -  & - &  - & - & -  & - &  -0.362 & 0.075 & 0.810 & 0.162 & -0.458 & 0.049 & -0.084 & 0.052 \\ 
 DIC   &   & \multicolumn{4}{c}{69401}   & \multicolumn{4}{|c}{68978}    & \multicolumn{4}{|c}{  68806 }   & \multicolumn{4}{|c}{68136} \\ 
 WAIC    &   & \multicolumn{4}{c}{69485}   & \multicolumn{4}{|c}{69067}    & \multicolumn{4}{|c}{68919}   & \multicolumn{4}{|c}{68281} \\ 
 Computational time    &    & \multicolumn{4}{c}{0.084 (0.012)}   & \multicolumn{4}{|c}{0.111 (0.020)}    & \multicolumn{4}{|c}{0.322 (0.075)}   & \multicolumn{4}{|c}{0.422 (0.159)} \\\hline
     \end{tabular}
\end{sidewaystable}
%\end{landscape}
\FloatBarrier

\subsection{Scenario 3}
The purpose of this scenario is to evaluate the performance and results of Model IV when implemented using two different Bayesian inference methods: Gibbs sampling through \texttt{R2OpenBUGS} and approximate Bayesian inference via \texttt{INLA}. Utilizing a consistent dataset (generated as in Scenario 2 with \( n_k = 20 \)), this comparison allows for an assessment of differences in estimation accuracy, computational efficiency, and overall feasibility between the two approaches. For the Gibbs sampler, we executed 10,000 iterations and assessed parameter convergence using the Gelman-Rubin criterion. The results, reported in Table \ref{sim3} and illustrated in Figure \ref{sen3}, span 100 simulations and present parameter estimates, standard errors, relative bias, and RMSE for various parameters. 
Additionally, the table and figure highlight both the accuracy and variability of the estimates across the two methods.  Notably, approximate Bayesian inference via \texttt{INLA} generally achieved closer estimates with lower computational demands, completing simulations in an average of 0.124 minutes compared to 339 minutes for Gibbs sampling. These results indicate that while both methods offer robust estimates, \texttt{INLA} demonstrates higher computational efficiency, making it potentially more feasible for large-scale or time-sensitive applications. The insights gained from this scenario provide valuable information on the trade-offs between Gibbs sampling and approximate Bayesian inference, particularly when balancing computational speed with estimation accuracy in spatial joint models.

\begin{figure}[hbt!]
\centering
\includegraphics[width=12cm]{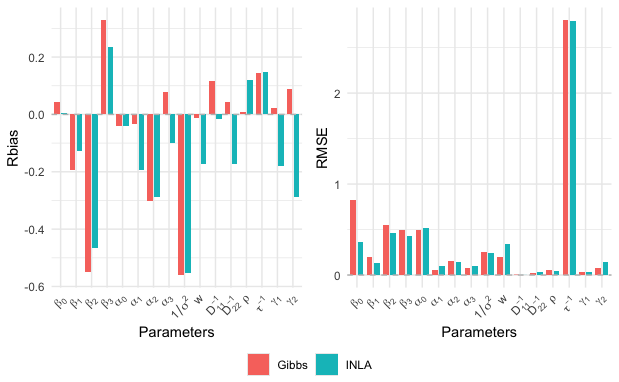}
\vspace*{-.1cm} \caption{\label{sen3}  
Comparison of Rbias and RMSE  between Gibbs Sampling and Approximate Bayesian Inference (INLA) methods under Scenario 3 of the simulation study. 
 }
\end{figure}
\FloatBarrier

\begin{table}
 \centering
 \tiny
  \caption{\label{sim3} Simulation Study Results for the Spatial Joint Model: Parameter Estimates, Standard Errors, Relative Bias, and RMSE for Scenario 3 with M=100 Simulations and $n_k=20$. The computational time is measured in minutes and includes the mean (standard deviation).
  }
  \begin{tabular}{c|c|cccc|cccc}
     \hline
                &   & \multicolumn{4}{c}{Approximate Bayesian Inference}   & \multicolumn{4}{|c}{Gibbs Sampling}     \\\hline
 Parameter &\ Real  &\ Est.  &\ S.E.  &\ Rbias &\ RMSE &\ Est.  &\ S.E.  &\ Rbias &\ RMSE \\\hline
$\beta_0$ & 9.000 & 9.045 & 0.519 & 0.005 & 0.367 & 9.400 & 1.162 & 0.044 & 0.822 \\ 
  $\beta_1$ & -1.000 & -0.872 & 0.066 & -0.128 & 0.128 & -0.805 & 0.132 & -0.195 & 0.195 \\ 
$\beta_2$ & -1.000 & -0.534 & 0.520 & -0.465 & 0.465 & -0.449 & 0.504 & -0.551 & 0.551 \\ 
$\beta_3$ & -1.500 & -1.854 & 0.607 & 0.236 & 0.429 & -1.994 & 0.507 & 0.329 & 0.494 \\ 
$\alpha_0$ & -6.000 & -5.758 & 0.424 & -0.040 & 0.519 & -5.755 & 0.698 & -0.041 & 0.494 \\ 
$\alpha_1$ & 0.500 & 0.404 & 0.049 & -0.193 & 0.096 & 0.483 & 0.074 & -0.034 & 0.052 \\ 
$\alpha_2$ & 0.500 & 0.356 & 0.178 & -0.288 & 0.144 & 0.349 & 0.210 & -0.303 & 0.151 \\ 
$\alpha_3$ & 1.000 & 0.901 & 0.011 & -0.099 & 0.099 & 1.078 & 0.038 & 0.078 & 0.078 \\ 
$1/\sigma^{2}$ & 0.447 & 0.200 & 0.006 & -0.552 & 0.247 & 0.197 & 0.001 & -0.560 & 0.251 \\ 
 $\varpi$ & 2.000 & 1.654 & 0.136 & -0.173 & 0.346 & 1.975 & 0.277 & -0.012 & 0.196 \\ 
$D_{11}^{-1}$ & 0.040 & 0.039 & 0.001 & -0.017 & 0.001 & 0.045 & 0.001 & 0.118 & 0.005 \\ 
$D_{22}^{-1}$ & 0.167 & 0.138 & 0.030 & -0.172 & 0.029 & 0.174 & 0.039 & 0.044 & 0.027 \\ 
 $\rho$ & -0.327 & -0.365 & 0.050 & 0.119 & 0.039 & -0.330 & 0.070 & 0.010 & 0.050 \\ 
$\tau^{-1}$ & 12.000 & 13.791 & 2.885 & 0.149 & 2.789 & 13.731 & 2.901 & 0.144 & 2.803 \\ 
$\gamma_1$ & -0.200 & -0.164 & 0.014 & -0.181 & 0.036 & -0.205 & 0.041 & 0.024 & 0.029 \\ 
$\gamma_2$ & -0.500 & -0.355 & 0.058 & -0.289 & 0.145 & -0.545 & 0.113 & 0.089 & 0.080 \\ 
   \hline
   Computational time    &    & \multicolumn{4}{c}{0.124 (0.075)}   & \multicolumn{4}{|c}{339 (12.543) }  \\\hline
\end{tabular}
\end{table}
\FloatBarrier

\subsection{Key findings from simulation studies}
In summary, the simulation study demonstrates that the proposed spatial joint model with spatial random effects (Model IV) consistently outperforms the other models across various parameter estimation metrics. Model IV effectively incorporates spatial correlations and jointly analyzes outcomes, resulting in significantly lower bias and RMSE values compared to separate models (Model I and II) and the joint model without spatial effects (Model III). The findings highlight the critical importance of considering spatial effects and employing a joint modeling approach for accurate parameter estimates. Additionally, larger sample sizes enhance estimation precision, underscoring the need for adequate data in spatial joint modeling contexts. Model IV is further validated by DIC and WAIC metrics, confirming its superior fit to the data and establishing it as the optimal choice for the analyzed scenarios.
}

\section{Discussion }
This paper presents an innovative approach to address the joint modeling of longitudinal measurements and spatial survival data, building upon the groundwork laid by \cite{martins2016bayesian} through the utilization of a Bayesian framework. While Bayesian methods offer powerful tools for inference in complex models, the computational demands associated with Markov Chain Monte Carlo (MCMC) algorithms can be prohibitive, especially in the context of joint modeling where multiple data sources are integrated. The computational complexity and time-consuming nature of MCMC hinder its scalability and practical application, particularly in large-scale studies or scenarios where timely analyses are imperative.\\
In response to these challenges, the study proposes an alternative methodology that leverages the Integrated Nested Laplace Approximation (\texttt{INLA}) technique. \texttt{INLA} provides a computationally efficient approximation method for Bayesian inference, offering considerable reductions in computation time compared to traditional MCMC approaches. By circumventing the need for lengthy MCMC iterations, \texttt{INLA} enables researchers to obtain accurate estimates and credible intervals in a fraction of the time, facilitating the rapid analysis of longitudinal and spatial data. This advancement is particularly significant in fields such as epidemiology and public health, where timely insights are crucial for informing policy and intervention strategies.\\
Furthermore, the methodology introduced in this paper addresses an important aspect of longitudinal data analysis—the nonlinear relationship between observed time and longitudinal responses. Traditional linear models may inadequately capture the complex temporal dynamics inherent in longitudinal data, leading to biased estimates and diminished model fit. By incorporating spline functions, the proposed approach offers a flexible framework capable of accommodating nonlinear trends in longitudinal responses. This enhancement not only improves the model's predictive accuracy but also enhances its interpretability by capturing the nuanced relationships between time and outcome variables.\\
{Overfitting occurs when a model becomes too complex, capturing both the true underlying patterns in the data and random noise, which results in poor generalization to unseen data \cite{hastie2009elements}. In our paper, we address this concern by employing several techniques that mitigate overfitting. First, our Bayesian hierarchical framework incorporates prior distributions that regularize the model, preventing it from becoming overly complex \cite{gelman2014bayesian}. We also use penalized splines for modeling the non-linear time effects, where the optimal number of knots is selected based on generalized cross-validation to avoid unnecessary flexibility \cite{eilers1996flexible}. Furthermore, we model spatial dependencies using a GMRF, which includes precision matrices that regularize the spatial effects and reduce the risk of overfitting \cite{rue2009approximate}.\\
To assess and quantify overfitting, we apply model selection criteria such as the Deviance Information Criterion (DIC) and Watanabe-Akaike Information Criterion (WAIC), both of which penalize models based on their complexity and ensure a balance between goodness of fit and simplicity \cite{spiegelhalter2002bayesian,gelman2014bayesian}. Additionally, we evaluate relative bias (Rbias) and root of mean squared error (RMSE) in our simulation studies to ensure that our models are accurate and robust without overfitting. The consistency of these metrics across different simulations, including various sample sizes and model complexities, demonstrates that our approach effectively generalizes to unseen data without overfitting.}
The provision of R code for implementing the proposed spatial joint model enhances the accessibility and reproducibility of the research findings. The R code is available in \url{https://github.com/tbaghfalaki/ASJM}.
Researchers can readily replicate the analyses conducted in the study, explore alternative modeling strategies, and adapt the methodology to suit diverse research contexts and datasets.\\
Looking forward, there are several promising avenues for extending the proposed framework. One potential direction involves the expansion of the model to accommodate multivariate longitudinal data or multivariate mixed longitudinal data, where multiple correlated outcomes are observed over time. By incorporating additional response variables, researchers can gain deeper insights into complex biological processes and disease trajectories, thereby enhancing the model's utility in clinical and epidemiological research.\\
Additionally, addressing the challenges associated with competing risks represents another important area of future investigation. Competing risks arise when individuals face multiple mutually exclusive events, such as death from different causes, and accounting for these complexities is essential for accurate risk estimation and decision-making. Developing robust methodologies for handling competing risks within the framework of joint modeling will further enhance the model's applicability in diverse research settings.\\
{Furthermore, addressing missing data remains a critical area of ongoing research, particularly in longitudinal studies where dropout, intermittent non-response, or measurement error is common. Proper handling of missing data is crucial for ensuring unbiased inference and reliable results. By incorporating advanced imputation methods or adopting principled approaches such as multiple imputation, researchers can effectively mitigate the effects of missing data on model estimates. Specifically, for spatial data, the authors recommend leveraging the inherent spatial structure to impute missing values based on neighboring regions, as the Besag model assumes spatial dependence \cite{besag1974spatial}. A simple approach involves imputing missing values using the mean of neighboring regions, while a more sophisticated option is model-based imputation using Bayesian inference, where missing values are estimated as part of the overall model through techniques such as MCMC \cite{rue2009approximate}. Additionally, the proper Besag model with regularization can stabilize the imputation process by borrowing strength from observed data. To further account for uncertainty in missing data, multiple imputation techniques can be applied, combining results from multiple imputed datasets to enhance the robustness of findings \cite{rubin1987multiple}.\\
This paper primarily focuses on right censoring of survival data. Although other types of censoring, such as interval or left censoring, are not explicitly addressed, the proposed model could potentially be extended to accommodate these mechanisms. Adapting the model for different types of censoring would likely necessitate some modifications; however, the general framework may still be applicable with the appropriate adjustments.}\\
In summary, the methodology proposed in this paper represents a significant advancement in the field of joint modeling, offering a versatile and efficient framework for analyzing complex longitudinal and spatial data. By combining innovative computational techniques with sophisticated modeling strategies, this research opens new avenues for exploring the intricate relationships between longitudinal measurements and spatial survival outcomes, ultimately contributing to improved understanding and management of disease processes and public health challenges.\\
{For future work, the proposed model could be extended to handle zero-inflated longitudinal outcomes \cite{baghfalaki2021approximate}, addressing scenarios where excess zeros occur alongside spatial dependencies for more comprehensive real-world applications.}

%\section*{Acknowledgements}
%We want to thank\ldots

{\footnotesize{
\bibliographystyle{plain}
\bibliography{wileyNJD-AMA.bib}}}

\end{document}